\documentclass[aps,prl,reprint,superscriptaddress,longbibliography]{revtex4-1}

\usepackage{braket}
\usepackage{amsmath}
\usepackage[textsize=tiny]{todonotes}
\DeclareMathOperator{\sgn}{sgn}

\begin{document}

\title{Topological phase transitions in the repulsively interacting Haldane-Hubbard model}

\author{Tuomas I. Vanhala}
\affiliation{COMP Centre of Excellence, Department of Applied Physics, Aalto University, Helsinki, Finland}
\affiliation{Theoretische Physik, ETH Zurich, 8093 Zurich, Switzerland}
\author{Topi Siro}
\author{Long Liang}
\affiliation{COMP Centre of Excellence, Department of Applied Physics, Aalto University, Helsinki, Finland}
\author{Matthias Troyer}
\affiliation{Theoretische Physik, ETH Zurich, 8093 Zurich, Switzerland}
\author{Ari Harju}
\email{ari.harju@aalto.fi}
\affiliation{COMP Centre of Excellence, Department of Applied Physics, Aalto University, Helsinki, Finland}
\author{P\"{a}ivi T\"{o}rm\"{a}}
\email{paivi.torma@aalto.fi}
\affiliation{COMP Centre of Excellence, Department of Applied Physics, Aalto University, Helsinki, Finland}
\affiliation{Institute for Quantum Electronics, ETH Zurich, 8093 Zurich, Switzerland}

\date{\today}

\begin{abstract}
Using dynamical mean-field theory and exact diagonalization we study the phase diagram of the repulsive Haldane-Hubbard model, varying the interaction strength and the sublattice potential difference. In addition to the quantum Hall phase with Chern number $C=2$ and the band insulator with $C=0$ present already in the noninteracting model, the system also exhibits a $C=0$ Mott insulating phase, and a $C=1$ quantum Hall phase. We explain the latter phase by a spontaneous symmetry breaking where one of the spin-components is in the Hall state and the other in the band insulating state.
\end{abstract}

\maketitle

When a quantum system has two or more competing phases, exotic states can emerge in the crossover region between these. Especially interesting phenomena can be expected between topologically trivial and non-trivial phases. Three paradigm models that offer a generic platform to explore such intermediate phases are the Haldane-Hubbard, Kane-Mele-Hubbard and the ionic Hubbard models. The existence and nature of exotic intermediate states of matter between such phases, showing spectral features and responses of a mixed character, is a subtle and largely open question.

In the ionic Hubbard models an energy offset (staggering) $\Delta_{AB}$ between the two sites ($A$ and $B$) of a bipartite lattice is combined with an on-site repulsive interaction $U$.
Starting from $\Delta_{AB}=U=0$, the models show a band insulator for large $\Delta_{AB}$ and a Mott insulator for strong interactions $U$.
Predictions of possible intermediate states between the the two insulators range from semimetals \cite{Ebrahimkhas20151053} and half-metals \cite{PhysRevB.91.235108} to metallic \cite{PhysRevLett.98.046403,PhysRevLett.97.046403,PhysRevB.79.121103} and insulating \cite{PhysRevLett.98.016402,PhysRevB.79.121103} ones, depending on the kinetic part of the Hamiltonian. Dimerized bond-ordered insulators have been shown to exist for 1D systems \cite{PhysRevLett.83.2014,PhysRevB.63.235108,PhysRevLett.92.246405}, but predictions for e.g. the 2D square lattice are contradictory \cite{PhysRevLett.98.046403,PhysRevLett.98.016402}. The band and Mott insulators in the ionic Hubbard model have been recently observed in ultracold quantum gases \cite{PhysRevLett.115.115303}, but no information about a possible intermediate state was obtained.

\begin{figure}
\includegraphics[width=0.7\columnwidth]{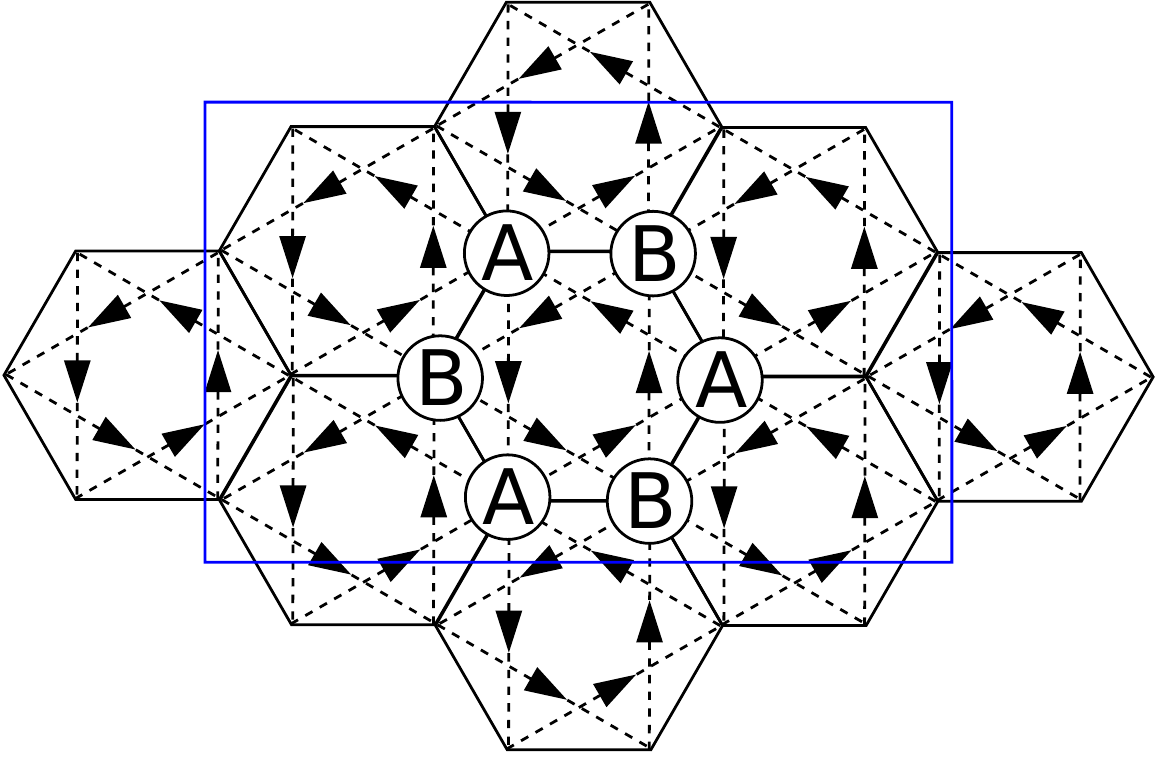}
\caption{A patch of the Haldane model. The model consists of a hexagonal lattice with nearest- and next-nearest-neighbour (nnn) hoppings. The arrows show the direction of positive phase winding for the complex nnn hoppings, which are responsible for the topological properties of the model. In this Letter we study the interplay of a potential difference between sublattices A and B and a local Hubbard interaction. The blue rectangle shows the finite size exact diagonalization cluster.\label{LatticeFigure}}
\end{figure}

In the Haldane model \cite{PhysRevLett.61.2015}, a staggered magnetic flux threads a hexagonal lattice, endowing the noninteracting electronic bands with a finite Chern number 
and quantized Hall conductivity. Large staggering $\Delta_{AB}$ drives the system from this topological 
insulator into a trivial band insulator. The model was recently realized in ultracold gas implementations \cite{2014Natur.515..237J,2015arXiv150905763F}. Large $U$ leads to a topologically trivial Mott insulator phase, but little is known about possible intermediate states. 
Mean-field studies \cite{PhysRevB.84.035127,2012PhRvB..85t5107H,0953-8984-26-17-175601,Wu2015,2016PhyB..481...53P} suggest existence of an interesting insulator phase with $C=1$, but whether this phase survives the inclusion of correlations is so far unknown. The existence of intermediate states is an open question \cite{PhysRevB.89.235104,zong2013quantum,2010NatPh...6..376P,
PhysRevB.85.115132,PhysRevB.91.045122} also in the similar but time-reversal-symmetric Kane-Mele model \cite{PhysRevLett.95.226801}, despite the fact that sign problem free quantum Monte Carlo (QMC) methods exist for that model.

In this Letter, we ask whether intermediate phases are possible in the Haldane-Hubbard model when the staggering $\Delta_{AB}$ and interaction $U$ are varied from zero to large values. We aim to investigate the nature of such states as well as their spectral properties, and to suggest feasible experimental realizations of the predicted phases. Importantly, mean-field theory is expected to be highly unreliable for the intermediate phases, not only due to strong interactions and low dimensionality (2D), but also because they are by definition states where orders of the surrounding phases compete. Therefore, a crucial ingredient of our study is that we apply {\it two complementary}, state-of-the-art beyond-mean-field methods. First, we perform exact diagonalization of finite-size clusters (FS-ED). Exact diagonalization gives reliable information about the nature of the ground state, without any bias from an ansatz, and proves its stability against quantum fluctuations over the system size, but suffers from finite size effects. Therefore we also apply dynamical mean-field theory (DMFT) to prove that the predicted phase survives at the thermodynamic limit. DMFT goes beyond static mean-field (MF) theory by treating local quantum fluctuations exactly. Non-local quantum fluctuations are not included and the method might be biased by the choice of the order parameters. In our case, however, these weaknesses are controlled by the exact diagonalization results. For comparison, we also present MF results.

We write the Hamiltonian of the Haldane-Hubbard model (Fig.\ \ref{LatticeFigure}) as $H=H_k+H_l$, where $H_l$ is a local, on-site part and the kinetic term $H_k$ is given by
\begin{equation}
H_k =t\sum_{\langle i,j \rangle,\sigma} c_{i\sigma}^{\dagger} c_{j\sigma} + t' \sum_{\langle\langle i,j \rangle\rangle,\sigma} \exp(i\phi_{ij})c_{i\sigma}^{\dagger} c_{j\sigma} 
\end{equation}
where $\left<i,j\right>$ and $\langle\langle i,j \rangle\rangle$ denote summation over nearest and next-nearest neighbours on a hexagonal lattice,  and $\sigma$ runs over the two spin components. The phase $\phi_{ij}$ has a constant absolute value and a sign that depends on the direction of the bond, $\phi_{ij}=\pm \phi$. The on-site part can be written as
\begin{equation}
H_l=U\sum_i \left(n_{i\uparrow}-\frac{1}{2} \right) \left(n_{i\downarrow}-\frac{1}{2} \right) +\Delta_{AB}\sum_{i,\sigma} \sgn(i)n_{i\sigma},
\end{equation}
where $\sgn(i)$ is $+1$ for sites $i$ on sublattice $A$ and $-1$ for sublattice $B$. In the following we take $t=1$ and set $t'=0.2$. A particle-hole transformation $c_{i\sigma} ' = \sgn(i) c_{i\sigma}^\dagger$ keeps the Hamiltonian otherwise invariant, but takes $\Delta_{AB}'=-\Delta_{AB}$ and $\phi'=\pi-\phi$. A further rotation  of the lattice by $\pi$ radians around a center of a hexagon only changes the sign of $\Delta_{AB}$. We study the model at half-filling and set the phase $\phi=\pi/2$. The above symmetries then imply that the phase diagram is symmetric under the reflection $\Delta_{AB} \rightarrow -\Delta_{AB}$ and that the chemical potential is zero. We do not consider large values of $t'$ for which chiral spin liquid phases, topological Mott insulators and exotic kinds of magnetic order have been proposed to appear 
\cite{1367-2630-18-3-035004,2015PhRvB..91m4414H,2015arXiv150908461H,PhysRevB.91.161107,2015arXiv151205118G,2015arXiv151204498W,PhysRevB.84.035127}, or attractive interactions which may support topological superfluids \cite{2015arXiv151103833Z,Wu2015}.

We study the phase diagram of this model as a function of the interaction strength $U$ and the sublattice potential difference $\Delta_{AB}$. We use an exact diagonalization impurity solver \cite{RevModPhys.68.,PhysRevLett.72.1545} to obtain results within single-site and two-site cellular DMFT \cite{RevModPhys.68.,RevModPhys.77.,PhysRevLett.87.186401}, always allowing for a symmetry breaking between the $A$ and $B$ sublattices. We find that using $5$ or $6$ bath sites already gives a good representation of the bath Green's function.  For selected parameters we have confirmed the results using the CT-INT algorithm \cite{PhysRevB.72.035122,ContinuousTime} as the impurity solver. In the mean-field and DMFT solutions antiferromagnetism is measured by an order parameter $m$ defined as
\begin{equation}
m=\frac{1}{N_s}\left|\sum_i  \sgn(i) ( \left\langle n_{i\uparrow} \right\rangle - \left\langle n_{i\downarrow}  \right\rangle) \right|,
\end{equation}
where $N_s$ is the number of sites, and the degree to which the particles are localized to the low-energy sublattice is measured by the staggered density
\begin{equation}
n_s=\frac{1}{N_s} \left| \sum_i  \sgn(i) (  \left\langle n_{i\uparrow} \right\rangle + \left\langle n_{i\downarrow} \right\rangle) \right|.
\end{equation}

Another important quantity is the Chern number. In \cite{PhysRevX.2.031008} it was shown, that knowledge of the zero-frequency Green's function is sufficient to determine topological invariants for interacting systems. The result can be formulated \cite{0953_8984_25_15_155601,PhysRevLett.113.136402} by defining the so-called topological hamiltonian as
\begin{equation}
h_t(\vec{k}) \equiv  -G(i\omega=0,\vec{k})^{-1} = h_0(\vec{k})+\Sigma(i\omega=0,\vec{k}),
\end{equation}
where $G$ and $\Sigma$ are the interacting single-particle Green's function and self energy of the problem, and $h_0$ is the noninteracting single-particle hamiltonian (the Bloch hamiltonian). In our case $G$, $\Sigma$ and $h_0$ are matrices in spin and sublattice space. Accoring to the theory, the Chern number calculated for a noninteracting problem defined by $h_t$ is the same as the Chern number for the original interacting problem. In DMFT we obtain an approximation for the zero-frequency self-energy by a linear interpolation between the smallest-in-absolute-value Matsubara frequencies at a very low temperature. We take care that the Matsubara frequency grid is dense enough to give an essentially smooth self energy near zero frequency and then use the topological Hamiltonian to calculate the Chern number using the method presented in \cite{FukuiTakahiroHatsugai2005}.

The Chern number can also be calculated from FS-ED using twisted boundary conditions \cite{PhysRevB.31.3372},
\begin{equation}
\psi(x_j+L_j) = e^{i\theta_j}\psi(x_j),
\end{equation}
 where $j$ indexes the space dimensions and $L_j$ is the length of the system along direction $j$. The Chern number can then be calculated by dividing the $(\theta_1,\theta_2)$ plane into a discrete lattice and computing the flux of the Berry curvature from the Berry phase acquired by the state around each cell \cite{Resta_2011}. Our results have been obtained for the $16$-site cluster shown in Fig. \ref{LatticeFigure}.

\begin{figure}[t]
        \includegraphics[width=0.9\columnwidth]{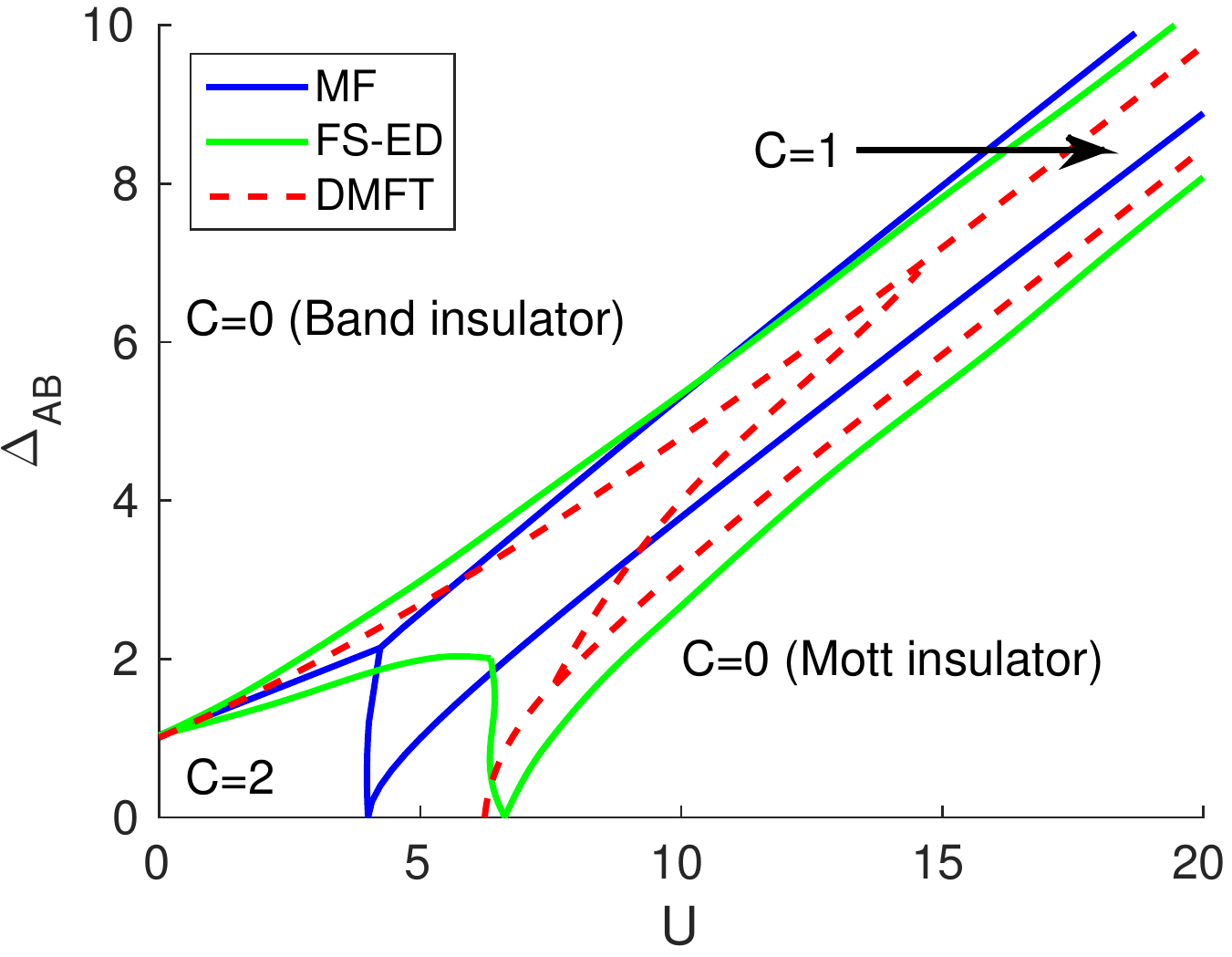}
        \caption{The phase diagram of the model from mean-field theory (MF), finite size exact diagonalization (FS-ED) and single-site dynamical mean-field theory (DMFT). The lines indicate the topological transitions where the Chern number $C$ changes. The most interesting feature is the $C=1$ phase found by all methods between the Mott insulating and band insulating regions.\label{PhaseDiagramFigure}}
\end{figure}

Our main result, the topological phase diagram of the model, is presented in Fig.\ \ref{PhaseDiagramFigure}. For small $U$ the main effect of the interaction is to push the transition from the quantum Hall phase to the band insulating phase to higher values of $\Delta_{AB}$ than in the noninteracting case. This effect can be explained in a mean-field picture: The sublattice potential difference causes a density difference between the sublattices, which in turn causes a Hartree potential that opposes this effect. The QH phase has Chern number $C=2$, which is the sum of the Chern numbers of the two (equivalent) spin channels. 

For high interaction strengths, on the other hand, the main features of the phase diagram are the antiferromagnetic Mott insulator and the band insulator phase whose boundary roughly follows the line $\Delta_{AB}=U/2$. Below this line $U$ dominates and the ground state of the local Hamiltonian $H_l$ has no doubly occupied sites, while above the line $\Delta_{AB}$ drives all of the particles to the lower energy sublattice. Nontrivial competition between the hopping, $U$ and $\Delta_{AB}$ occurs close to the line $\Delta_{AB}=U/2$ where the large energy scales $U$ and $\Delta_{AB}$ mostly cancel each other. Indeed, we find that this boundary region between the two topologically trivial insulators exhibits a phase with Chern number $C=1$.

In the mean-field solution \cite{PhysRevB.84.035127,2012PhRvB..85t5107H,0953-8984-26-17-175601,Wu2015,2016PhyB..481...53P} for the $C=1$ phase the staggered potential drives one of the components mostly to the low-energy sublattice. Thus, this component is effectively in the topologically trivial region of the phase diagram of the Haldane model. However, the larger density \emph{of one component} on the lower sublattice creates a Hartree potential that mostly cancels the sublattice potential difference $\Delta_{AB}$ \emph{for the other component}, which then carries the Chern number $C=1$. Because of this symmetry breaking, the $C=1$ phase has a nonzero $m$, while the $C=2$ phase and the band insulator are paramagnetic, although the mean-field solution also has a very narrow antiferromagnetic region with $C=2$ near the $C=2$ to $C=1$ transition line.

\begin{figure}[t]
\includegraphics[width=1\columnwidth]{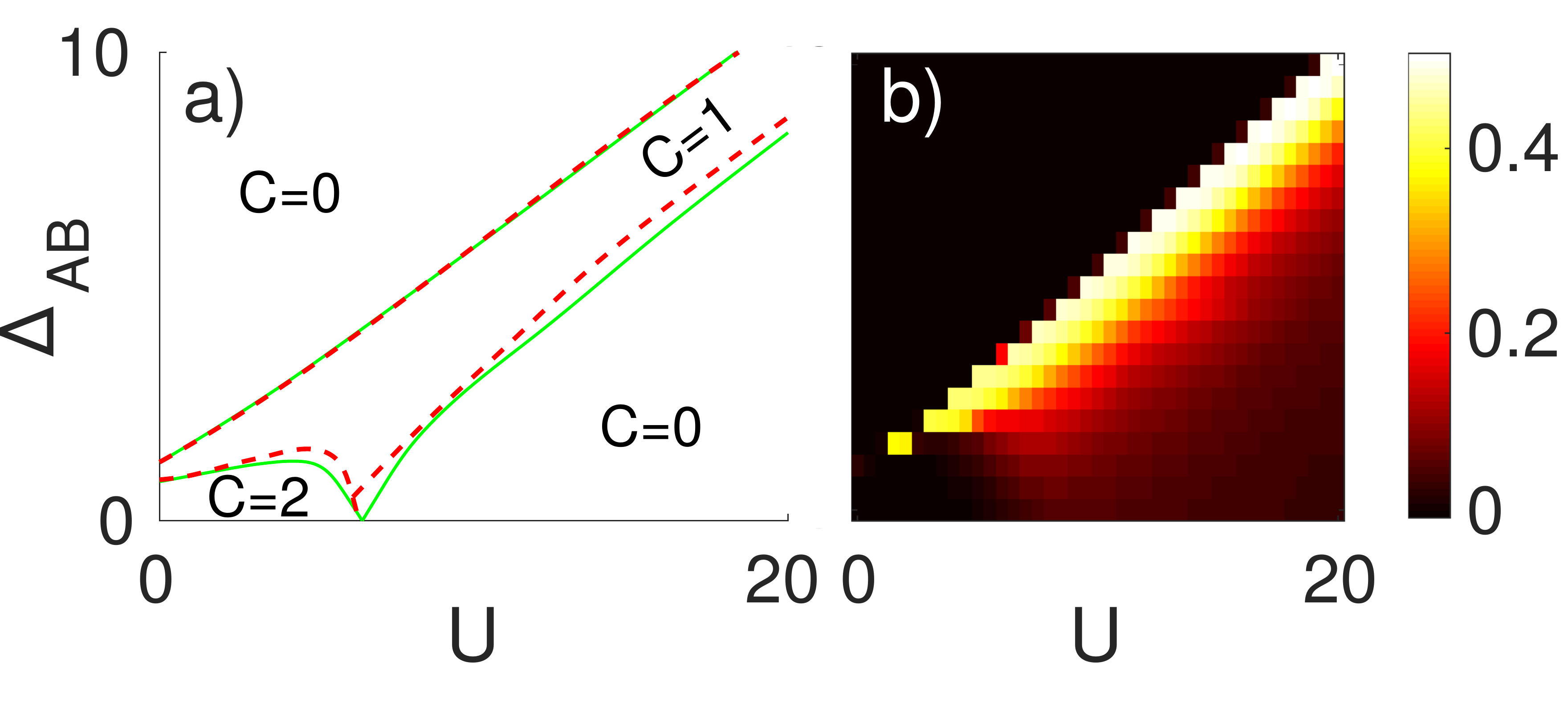}
\caption{a) The single-site DMFT and FS-ED phase diagrams for a finite mass imbalance. The hoppings for the up-component were scaled by a factor of $0.8$, while the hoppings for the down-component were scaled by $1.2$ compared to the balanced situation in Fig.\ \ref{PhaseDiagramFigure}.\ b) The overlap between the ground state and the trial state $\ket{\psi}$ obtained from the FS-ED calculations (see text) for the same parameters as in Fig.\ \ref{PhaseDiagramFigure}.\label{CombinedFigure}}
\end{figure}%

We have confirmed this picture in the FS-ED calculations by comparing the obtained ground state with an ansatz that is a symmetric linear combination $\ket{\psi}=\frac{1}{\sqrt{2}} \left( \ket{\text{QH}}_{\uparrow} \ket{\text{BI}}_{\downarrow} + \ket{\text{BI}}_{\uparrow} \ket{\text{QH}}_{\downarrow} \right)$, where $\ket{\text{QH}}$ and $\ket{\text{BI}}$ are the single-component ground states of the noninteracting model for vanishing and large $\Delta_{AB}$, respectively. In Fig.\ \ref{CombinedFigure}b we present the overlap between this state and the ground state, which reaches values as high as $0.5$ in the $C=1$ region of the phase diagram. This shows that the above qualitative picture of the $C=1$ state is correct. In the FS-ED results the $C=1$ phase is present already for weak interactions. However, this is a finite size effect: A mean field calculation for the FS-ED cluster produces the same result for weak $U$, while the $C=1$ phase is absent in the infinite-lattice mean-field in this region. This is expected, as finite size effects are known to be important when the band gap of the noninteracting Hamiltonian is small \cite{PhysRevB.84.241105}.

\begin{figure}[t]
\includegraphics[width=0.9\columnwidth]{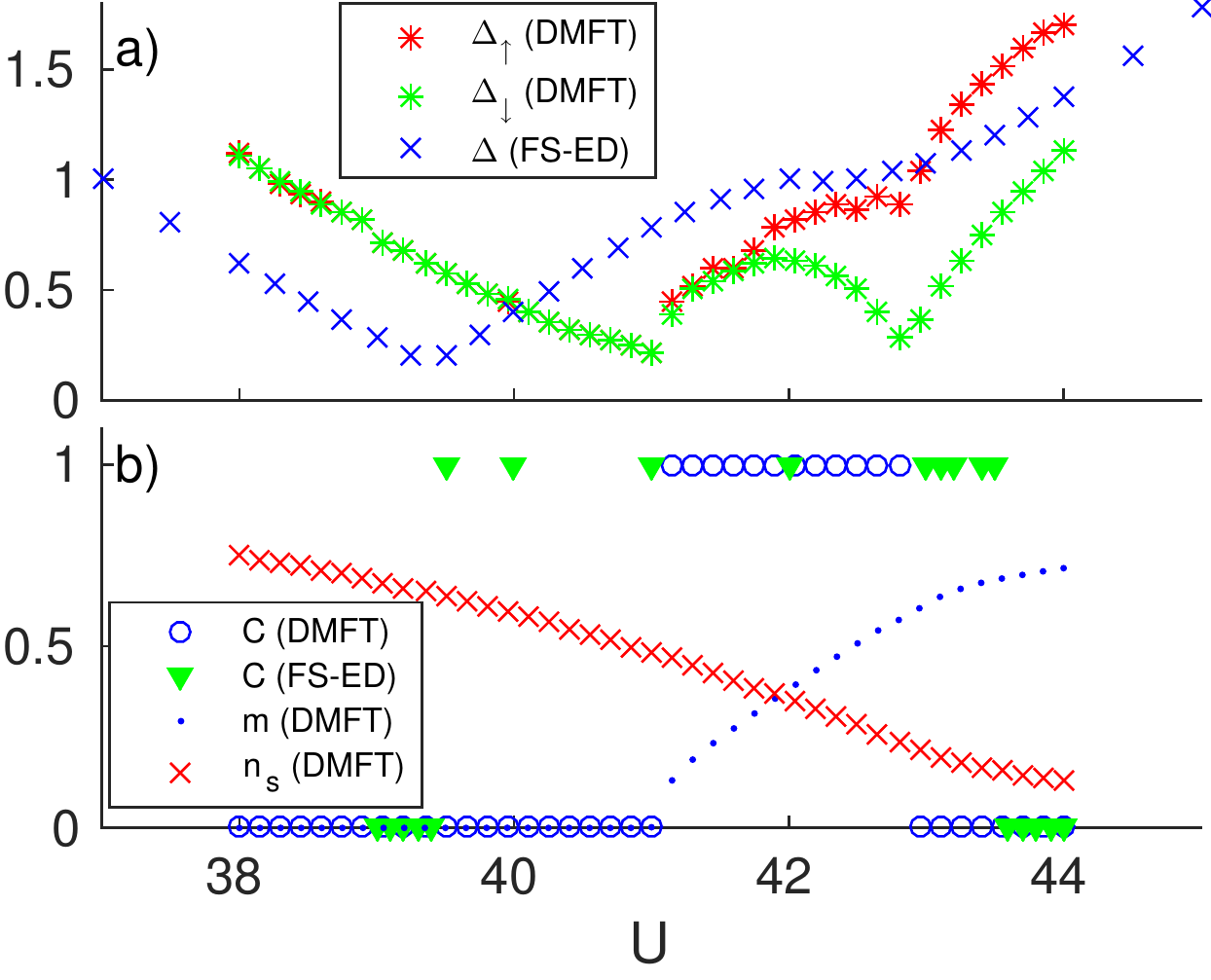}
\caption{a) The quasiparticle gap $\Delta$ for the up and down components obtained from two-site DMFT and the spin-rotation-invariant result from FS-ED at $\Delta_{AB}=20$. b) The Chern number, staggered density and the antiferromagnetic order parameter $m$ from two-site DMFT and FS-ED for the same $\Delta_{AB}$. \label{GapFigure}}
\end{figure}

To further understand the nature of the $C=1$ phase we have calculated the quasiparticle gap (see Fig.\ \ref{GapFigure}), which in ultracold gas experiments can be studied using for instance RF or lattice modulation spectroscopy \cite{PaivinSpektroskopia2015}. For $U \ll 2\Delta_{AB}$ the system is in the band insulating state and we find a gap that gets smaller as $U$ is increased. The gap has a minimum at the point where the system enters the $C=1$ state. When $U$ is increased furter, the gap for the component that carries Chern number $C=1$ again reaches a minimum, and the system moves to the $C=0$ Mott insulator phase, where we see a gap that grows as a function of $U$. In the FS-ED calculation we do not see a gap closing at the boundary of the $C=1$ phase and the Mott phase because the finite size ground state is symmetric with respect to spin rotations.

We have also done DMFT calculations for different values of $t'$. When $t'$ is increased, the intermediate band between the band insulator and the Mott insulator gets wider, as the Mott insulator is pushed to larger values of $U$. At the same time the $C=2$ phase extends into higher values of $U$ within the intermediate band: for $t'=0.3$ it is already present at values of $U>20$. However, we stress that our DMFT results for $t' \gtrsim 0.3$ are not necessarily physical, as we have not considered e.g. the exotic magnetic orders predicted for this parameter region \cite{2015arXiv150908461H,2015PhRvB..91m4414H}.

For $\Delta_{AB}=0$ we have performed six-site cluster DMFT calculations \cite{PhysRevB.87.205127} using the CT-INT impurity solver, as there is only a minor sign problem in this region. All DMFT results show a first order transition, which explains why FS-ED results differ from DMFT close to $\Delta_{AB}=0$: The finite cluster cannot exhibit sharp phase transitions where long range order suddenly develops. We note that for vanishing $t'$ the transition is indicated by large scale QMC to be of the second order \cite{2012NatSR...2E.992S}, while cluster DMFT finds a first order transition \cite{PhysRevB.86.045105}. Extrapolating to zero temperature, we find that the existence of the AF solution in DCA starts at $U_c=6.21 \pm 0.01$, which is in good agreement with the single-site DMFT result $U_c=6.27 \pm 0.01$, while cellular DMFT gives a slightly higher value $U_c=6.6 \pm 0.1$. Wu et al.\ \cite{2015arXiv151204498W} find a much lower value $U_c \approx 3.7$ at $t'=0.2$ in a two-site DMFT scheme. However, our results for $U_c$ agree well with the FS-ED transition point and are consistent with the $t'=0$ cellular DMFT result $U_c \approx 4.6$ \cite{PhysRevB.86.045105}, as $t'>0$ is expected to increase $U_c$.

Finally, we note that by the Mermin-Wagner theorem the continuous $SU(2)$ symmetry of the model can only be broken at zero temperature. However, the critical temperatures can be made finite by adding contributions to the Hamiltonian that break the $SU(2)$ symmetry \cite{PhysRevB.67.104414,PhysRevB.68.060402}. A mass imbalance (i.e. different hopping strengths for the two components) would even explicitly break the whole $SU(2)$ symmetry away from the $\Delta_{AB}=0$ line. As this enables the observation of a $C=1$ phase already for vanishing interactions, it would provide a way to test detection techniques in the noninteracting limit. Fig. \ref{CombinedFigure}a presents the DMFT phase diagram of the model for a finite mass imbalance, showing that the essential features of the balanced case are preserved.

In summary, we found firm evidence, by two complementary beyond-mean-field methods, for a $C=1$ insulator in the Haldane-Hubbard model which spontaneously breaks the $SU(2)$ spin-rotation symmetry of the model.
Our results differ from the mean-field result \cite{PhysRevB.84.035127,2012PhRvB..85t5107H,0953-8984-26-17-175601,Wu2015,
2016PhyB..481...53P} which sets the $\Delta_{AB}=0$ boundary between the $C=2$ and Mott insulators at $U=4t$, and predicts the $C=1$ phase to be present for small $\Delta_{AB}$. In contrast, we find the boundary at $U=6t$ and that the $C=1$ phase is more likely to occur only for $\Delta_{AB} \gtrsim 2$ as predicted by DMFT (the prediction of this phase by FS-ED for small $\Delta_{AB}$ is likely a finite-size effect).
It is also not clear if the slave-spin theory of \cite{2016PhyB..481...53P} is an improvement over the mean-field treatment as it finds that the critical $U$ for antiferromagnetism at the $\Delta_{AB}=0$ line is decreased compared to the mean-field result, and there is an unphysical first order transition to the Mott phase with $n_s$ as the order parameter.
Furthermore, we do not see signs of spin liquid phases predicted to occur \cite{2012PhRvB..85t5107H,2011PhRvB..83t5116H} already for small $t'$.

In comparison to the ionic Hubbard models, the intermediate state we found is more robust and occupies a larger part of the phase diagram. For instance the intermediate insulator state in \cite{PhysRevLett.98.016402} vanishes around $U=11$ while ours continues. In \cite{PhysRevLett.97.046403,Ebrahimkhas20151053}, large $\Delta_{AB}$ suppresses the intermediate phase while in our case it helps to stabilize it. The half-metal found in \cite{PhysRevB.91.235108}, which resembles our $C=1$ state since only one spin component is gapped, exists in a tiny parameter regime compared to the large stability area we find. Thus it seems evident that the intermediate phase in the Haldane-Hubbard model, compared to the ionic Hubbard one, is {\it stabilized by topological effects}. In contrast to the semi-metal of ionic Hubbard model, the phase diagram starts from $\Delta_{AB}=U=0$ as a topological insulator with $C=2$. Characteritics of such an insulator, for one component, are inherited in the $C=1$ phase. From the comparison to the mean-field studies of the Haldane-Hubbard model and to the ionic Hubbard model results we conclude that while correlation effects tend to destroy the $C=1$ phase for small $\Delta_{AB}$, it survives as an exceptionally stable intermediate state close to the $\Delta_{AB}=U/2$ line in the very strongly interacting region.

Experimental observation of the predicted phases would be of fundamental importance for understanding not only the Haldane-Hubbard model but also the intermediate states in its cousin models. We calculate the quasiparticle gap, and suggest that it could be used for probing the phase diagam experimentally. Finally, considering the difficulty of achieving very low temperatures experimentally, we propose that a finite mass imbalance would make it easier to experimentally access the interesting features of the phase diagram.

\begin{acknowledgments}

We thank T. Esslinger, G. Jotzu, R. Desbuquois, M. Messer, and F. G\"{o}rg for useful discussions.
This  work  was  supported  by  the  Academy  of  Finland
through its Centers of Excellence Programme (2012-2017) and
under project Nos.\ 263347, 251748, 284621 and 272490, by the
European Research Council (ERC-2013-AdG-340748-CODE
and ERC-2011-AdG-290464-SIMCOFE), by the Swiss National Science Foundation through the National Competence Center in Research QSIT, and by the Pauli Center for Theoretical Studies at ETH Zurich.
 T.I.V. is grateful for
the support from the Vilho, Yrj\"{o} and Kalle V\"{a}is\"{a}l\"{a} Foundation.
T.S. acknowledges financial support from the Finnish
Doctoral Programme in Computational Sciences FICS.
Computing resources were provided by CSC -- the Finnish IT
Centre for Science and the Triton cluster at Aalto University.

\end{acknowledgments}

\bibliography{haldane_references}

\begin{thebibliography}{51}%
\makeatletter
\providecommand \@ifxundefined [1]{%
 \@ifx{#1\undefined}
}%
\providecommand \@ifnum [1]{%
 \ifnum #1\expandafter \@firstoftwo
 \else \expandafter \@secondoftwo
 \fi
}%
\providecommand \@ifx [1]{%
 \ifx #1\expandafter \@firstoftwo
 \else \expandafter \@secondoftwo
 \fi
}%
\providecommand \natexlab [1]{#1}%
\providecommand \enquote  [1]{``#1''}%
\providecommand \bibnamefont  [1]{#1}%
\providecommand \bibfnamefont [1]{#1}%
\providecommand \citenamefont [1]{#1}%
\providecommand \href@noop [0]{\@secondoftwo}%
\providecommand \href [0]{\begingroup \@sanitize@url \@href}%
\providecommand \@href[1]{\@@startlink{#1}\@@href}%
\providecommand \@@href[1]{\endgroup#1\@@endlink}%
\providecommand \@sanitize@url [0]{\catcode `\\12\catcode `\$12\catcode
  `\&12\catcode `\#12\catcode `\^12\catcode `\_12\catcode `\%12\relax}%
\providecommand \@@startlink[1]{}%
\providecommand \@@endlink[0]{}%
\providecommand \url  [0]{\begingroup\@sanitize@url \@url }%
\providecommand \@url [1]{\endgroup\@href {#1}{\urlprefix }}%
\providecommand \urlprefix  [0]{URL }%
\providecommand \Eprint [0]{\href }%
\providecommand \doibase [0]{http://dx.doi.org/}%
\providecommand \selectlanguage [0]{\@gobble}%
\providecommand \bibinfo  [0]{\@secondoftwo}%
\providecommand \bibfield  [0]{\@secondoftwo}%
\providecommand \translation [1]{[#1]}%
\providecommand \BibitemOpen [0]{}%
\providecommand \bibitemStop [0]{}%
\providecommand \bibitemNoStop [0]{.\EOS\space}%
\providecommand \EOS [0]{\spacefactor3000\relax}%
\providecommand \BibitemShut  [1]{\csname bibitem#1\endcsname}%
\let\auto@bib@innerbib\@empty
\bibitem [{\citenamefont {Ebrahimkhas}\ \emph {et~al.}(2015)\citenamefont
  {Ebrahimkhas}, \citenamefont {Drezhegrighash},\ and\ \citenamefont
  {Soltani}}]{Ebrahimkhas20151053}%
  \BibitemOpen
  \bibfield  {author} {\bibinfo {author} {\bibfnamefont {M.}~\bibnamefont
  {Ebrahimkhas}}, \bibinfo {author} {\bibfnamefont {Z.}~\bibnamefont
  {Drezhegrighash}}, \ and\ \bibinfo {author} {\bibfnamefont {E.}~\bibnamefont
  {Soltani}},\ }\bibfield  {title} {\enquote {\bibinfo {title} {Effects of
  correlations on honeycomb lattice in ionic-{H}ubbard model},}\ }\href
  {\doibase http://dx.doi.org/10.1016/j.physleta.2015.01.024} {\bibfield
  {journal} {\bibinfo  {journal} {Physics Letters A}\ }\textbf {\bibinfo
  {volume} {379}},\ \bibinfo {pages} {1053 -- 1056} (\bibinfo {year}
  {2015})}\BibitemShut {NoStop}%
\bibitem [{\citenamefont {Bag}\ \emph {et~al.}(2015)\citenamefont {Bag},
  \citenamefont {Garg},\ and\ \citenamefont
  {Krishnamurthy}}]{PhysRevB.91.235108}%
  \BibitemOpen
  \bibfield  {author} {\bibinfo {author} {\bibfnamefont {S.}~\bibnamefont
  {Bag}}, \bibinfo {author} {\bibfnamefont {A.}~\bibnamefont {Garg}}, \ and\
  \bibinfo {author} {\bibfnamefont {H.~R.}\ \bibnamefont {Krishnamurthy}},\
  }\bibfield  {title} {\enquote {\bibinfo {title} {Phase diagram of the
  half-filled ionic {H}ubbard model},}\ }\href {\doibase
  10.1103/PhysRevB.91.235108} {\bibfield  {journal} {\bibinfo  {journal} {Phys.
  Rev. B}\ }\textbf {\bibinfo {volume} {91}},\ \bibinfo {pages} {235108}
  (\bibinfo {year} {2015})}\BibitemShut {NoStop}%
\bibitem [{\citenamefont {Paris}\ \emph {et~al.}(2007)\citenamefont {Paris},
  \citenamefont {Bouadim}, \citenamefont {Hebert}, \citenamefont {Batrouni},\
  and\ \citenamefont {Scalettar}}]{PhysRevLett.98.046403}%
  \BibitemOpen
  \bibfield  {author} {\bibinfo {author} {\bibfnamefont {N.}~\bibnamefont
  {Paris}}, \bibinfo {author} {\bibfnamefont {K.}~\bibnamefont {Bouadim}},
  \bibinfo {author} {\bibfnamefont {F.}~\bibnamefont {Hebert}}, \bibinfo
  {author} {\bibfnamefont {G.~G.}\ \bibnamefont {Batrouni}}, \ and\ \bibinfo
  {author} {\bibfnamefont {R.~T.}\ \bibnamefont {Scalettar}},\ }\bibfield
  {title} {\enquote {\bibinfo {title} {Quantum {M}onte {C}arlo study of an
  interaction-driven band-insulator--to--metal transition},}\ }\href {\doibase
  10.1103/PhysRevLett.98.046403} {\bibfield  {journal} {\bibinfo  {journal}
  {Phys. Rev. Lett.}\ }\textbf {\bibinfo {volume} {98}},\ \bibinfo {pages}
  {046403} (\bibinfo {year} {2007})}\BibitemShut {NoStop}%
\bibitem [{\citenamefont {Garg}\ \emph {et~al.}(2006)\citenamefont {Garg},
  \citenamefont {Krishnamurthy},\ and\ \citenamefont
  {Randeria}}]{PhysRevLett.97.046403}%
  \BibitemOpen
  \bibfield  {author} {\bibinfo {author} {\bibfnamefont {A.}~\bibnamefont
  {Garg}}, \bibinfo {author} {\bibfnamefont {H.~R.}\ \bibnamefont
  {Krishnamurthy}}, \ and\ \bibinfo {author} {\bibfnamefont {M.}~\bibnamefont
  {Randeria}},\ }\bibfield  {title} {\enquote {\bibinfo {title} {Can
  correlations drive a band insulator metallic?}}\ }\href {\doibase
  10.1103/PhysRevLett.97.046403} {\bibfield  {journal} {\bibinfo  {journal}
  {Phys. Rev. Lett.}\ }\textbf {\bibinfo {volume} {97}},\ \bibinfo {pages}
  {046403} (\bibinfo {year} {2006})}\BibitemShut {NoStop}%
\bibitem [{\citenamefont {Byczuk}\ \emph {et~al.}(2009)\citenamefont {Byczuk},
  \citenamefont {Sekania}, \citenamefont {Hofstetter},\ and\ \citenamefont
  {Kampf}}]{PhysRevB.79.121103}%
  \BibitemOpen
  \bibfield  {author} {\bibinfo {author} {\bibfnamefont {K.}~\bibnamefont
  {Byczuk}}, \bibinfo {author} {\bibfnamefont {M.}~\bibnamefont {Sekania}},
  \bibinfo {author} {\bibfnamefont {W.}~\bibnamefont {Hofstetter}}, \ and\
  \bibinfo {author} {\bibfnamefont {A.~P.}\ \bibnamefont {Kampf}},\ }\bibfield
  {title} {\enquote {\bibinfo {title} {Insulating behavior with spin and charge
  order in the ionic {H}ubbard model},}\ }\href {\doibase
  10.1103/PhysRevB.79.121103} {\bibfield  {journal} {\bibinfo  {journal} {Phys.
  Rev. B}\ }\textbf {\bibinfo {volume} {79}},\ \bibinfo {pages} {121103}
  (\bibinfo {year} {2009})}\BibitemShut {NoStop}%
\bibitem [{\citenamefont {Kancharla}\ and\ \citenamefont
  {Dagotto}(2007)}]{PhysRevLett.98.016402}%
  \BibitemOpen
  \bibfield  {author} {\bibinfo {author} {\bibfnamefont {S.~S.}\ \bibnamefont
  {Kancharla}}\ and\ \bibinfo {author} {\bibfnamefont {E.}~\bibnamefont
  {Dagotto}},\ }\bibfield  {title} {\enquote {\bibinfo {title} {Correlated
  insulated phase suggests bond order between band and {M}ott insulators in two
  dimensions},}\ }\href {\doibase 10.1103/PhysRevLett.98.016402} {\bibfield
  {journal} {\bibinfo  {journal} {Phys. Rev. Lett.}\ }\textbf {\bibinfo
  {volume} {98}},\ \bibinfo {pages} {016402} (\bibinfo {year}
  {2007})}\BibitemShut {NoStop}%
\bibitem [{\citenamefont {Fabrizio}\ \emph {et~al.}(1999)\citenamefont
  {Fabrizio}, \citenamefont {Gogolin},\ and\ \citenamefont
  {Nersesyan}}]{PhysRevLett.83.2014}%
  \BibitemOpen
  \bibfield  {author} {\bibinfo {author} {\bibfnamefont {M.}~\bibnamefont
  {Fabrizio}}, \bibinfo {author} {\bibfnamefont {A.~O.}\ \bibnamefont
  {Gogolin}}, \ and\ \bibinfo {author} {\bibfnamefont {A.~A.}\ \bibnamefont
  {Nersesyan}},\ }\bibfield  {title} {\enquote {\bibinfo {title} {From band
  insulator to {M}ott insulator in one dimension},}\ }\href {\doibase
  10.1103/PhysRevLett.83.2014} {\bibfield  {journal} {\bibinfo  {journal}
  {Phys. Rev. Lett.}\ }\textbf {\bibinfo {volume} {83}},\ \bibinfo {pages}
  {2014--2017} (\bibinfo {year} {1999})}\BibitemShut {NoStop}%
\bibitem [{\citenamefont {Wilkens}\ and\ \citenamefont
  {Martin}(2001)}]{PhysRevB.63.235108}%
  \BibitemOpen
  \bibfield  {author} {\bibinfo {author} {\bibfnamefont {T.}~\bibnamefont
  {Wilkens}}\ and\ \bibinfo {author} {\bibfnamefont {R.~M.}\ \bibnamefont
  {Martin}},\ }\bibfield  {title} {\enquote {\bibinfo {title} {Quantum {M}onte
  {C}arlo study of the one-dimensional ionic {H}ubbard model},}\ }\href
  {\doibase 10.1103/PhysRevB.63.235108} {\bibfield  {journal} {\bibinfo
  {journal} {Phys. Rev. B}\ }\textbf {\bibinfo {volume} {63}},\ \bibinfo
  {pages} {235108} (\bibinfo {year} {2001})}\BibitemShut {NoStop}%
\bibitem [{\citenamefont {Batista}\ and\ \citenamefont
  {Aligia}(2004)}]{PhysRevLett.92.246405}%
  \BibitemOpen
  \bibfield  {author} {\bibinfo {author} {\bibfnamefont {C.~D.}\ \bibnamefont
  {Batista}}\ and\ \bibinfo {author} {\bibfnamefont {A.~A.}\ \bibnamefont
  {Aligia}},\ }\bibfield  {title} {\enquote {\bibinfo {title} {Exact bond
  ordered ground state for the transition between the band and the {M}ott
  insulator},}\ }\href {\doibase 10.1103/PhysRevLett.92.246405} {\bibfield
  {journal} {\bibinfo  {journal} {Phys. Rev. Lett.}\ }\textbf {\bibinfo
  {volume} {92}},\ \bibinfo {pages} {246405} (\bibinfo {year}
  {2004})}\BibitemShut {NoStop}%
\bibitem [{\citenamefont {Messer}\ \emph {et~al.}(2015)\citenamefont {Messer},
  \citenamefont {Desbuquois}, \citenamefont {Uehlinger}, \citenamefont {Jotzu},
  \citenamefont {Huber}, \citenamefont {Greif},\ and\ \citenamefont
  {Esslinger}}]{PhysRevLett.115.115303}%
  \BibitemOpen
  \bibfield  {author} {\bibinfo {author} {\bibfnamefont {M.}~\bibnamefont
  {Messer}}, \bibinfo {author} {\bibfnamefont {R.}~\bibnamefont {Desbuquois}},
  \bibinfo {author} {\bibfnamefont {T.}~\bibnamefont {Uehlinger}}, \bibinfo
  {author} {\bibfnamefont {G.}~\bibnamefont {Jotzu}}, \bibinfo {author}
  {\bibfnamefont {S.}~\bibnamefont {Huber}}, \bibinfo {author} {\bibfnamefont
  {D.}~\bibnamefont {Greif}}, \ and\ \bibinfo {author} {\bibfnamefont
  {T.}~\bibnamefont {Esslinger}},\ }\bibfield  {title} {\enquote {\bibinfo
  {title} {Exploring competing density order in the ionic {H}ubbard model with
  ultracold fermions},}\ }\href {\doibase 10.1103/PhysRevLett.115.115303}
  {\bibfield  {journal} {\bibinfo  {journal} {Phys. Rev. Lett.}\ }\textbf
  {\bibinfo {volume} {115}},\ \bibinfo {pages} {115303} (\bibinfo {year}
  {2015})}\BibitemShut {NoStop}%
\bibitem [{\citenamefont {{H}aldane}(1988)}]{PhysRevLett.61.2015}%
  \BibitemOpen
  \bibfield  {author} {\bibinfo {author} {\bibfnamefont {F.~D.~M.}\
  \bibnamefont {{H}aldane}},\ }\bibfield  {title} {\enquote {\bibinfo {title}
  {Model for a quantum hall effect without {L}andau levels: {C}ondensed-matter
  realization of the "parity anomaly"},}\ }\href {\doibase
  10.1103/PhysRevLett.61.2015} {\bibfield  {journal} {\bibinfo  {journal}
  {Phys. Rev. Lett.}\ }\textbf {\bibinfo {volume} {61}},\ \bibinfo {pages}
  {2015--2018} (\bibinfo {year} {1988})}\BibitemShut {NoStop}%
\bibitem [{\citenamefont {Jotzu}\ \emph {et~al.}(2014)\citenamefont {Jotzu},
  \citenamefont {Messer}, \citenamefont {Desbuquois}, \citenamefont {Lebrat},
  \citenamefont {Uehlinger}, \citenamefont {Greif},\ and\ \citenamefont
  {Esslinger}}]{2014Natur.515..237J}%
  \BibitemOpen
  \bibfield  {author} {\bibinfo {author} {\bibfnamefont {G.}~\bibnamefont
  {Jotzu}}, \bibinfo {author} {\bibfnamefont {M.}~\bibnamefont {Messer}},
  \bibinfo {author} {\bibfnamefont {R.}~\bibnamefont {Desbuquois}}, \bibinfo
  {author} {\bibfnamefont {M.}~\bibnamefont {Lebrat}}, \bibinfo {author}
  {\bibfnamefont {T.}~\bibnamefont {Uehlinger}}, \bibinfo {author}
  {\bibfnamefont {D.}~\bibnamefont {Greif}}, \ and\ \bibinfo {author}
  {\bibfnamefont {T.}~\bibnamefont {Esslinger}},\ }\bibfield  {title} {\enquote
  {\bibinfo {title} {Experimental realization of the topological {H}aldane
  model with ultracold fermions},}\ }\href {\doibase 10.1038/nature13915}
  {\bibfield  {journal} {\bibinfo  {journal} {\nat}\ }\textbf {\bibinfo
  {volume} {515}},\ \bibinfo {pages} {237--240} (\bibinfo {year}
  {2014})}\BibitemShut {NoStop}%
\bibitem [{\citenamefont {Fl{\"{a}}schner}\ \emph {et~al.}(2015)\citenamefont
  {Fl{\"{a}}schner}, \citenamefont {Rem}, \citenamefont {Tarnowski},
  \citenamefont {Vogel}, \citenamefont {L{\"{u}}hmann}, \citenamefont
  {Sengstock},\ and\ \citenamefont {Weitenberg}}]{2015arXiv150905763F}%
  \BibitemOpen
  \bibfield  {author} {\bibinfo {author} {\bibfnamefont {N.}~\bibnamefont
  {Fl{\"{a}}schner}}, \bibinfo {author} {\bibfnamefont {B.~S.}\ \bibnamefont
  {Rem}}, \bibinfo {author} {\bibfnamefont {M.}~\bibnamefont {Tarnowski}},
  \bibinfo {author} {\bibfnamefont {D.}~\bibnamefont {Vogel}}, \bibinfo
  {author} {\bibfnamefont {D.~S.}\ \bibnamefont {L{\"{u}}hmann}}, \bibinfo
  {author} {\bibfnamefont {K.}~\bibnamefont {Sengstock}}, \ and\ \bibinfo
  {author} {\bibfnamefont {C.}~\bibnamefont {Weitenberg}},\ }\bibfield  {title}
  {\enquote {\bibinfo {title} {Experimental reconstruction of the {B}erry
  curvature in a topological {B}loch band},}\ }\href@noop {} {\bibfield
  {journal} {\bibinfo  {journal} {ArXiv e-prints}\ } (\bibinfo {year}
  {2015})},\ \Eprint {http://arxiv.org/abs/1509.05763} {arXiv:1509.05763
  [cond-mat.quant-gas]} \BibitemShut {NoStop}%
\bibitem [{\citenamefont {He}\ \emph {et~al.}(2011)\citenamefont {He},
  \citenamefont {Zong}, \citenamefont {Kou}, \citenamefont {Liang},\ and\
  \citenamefont {Feng}}]{PhysRevB.84.035127}%
  \BibitemOpen
  \bibfield  {author} {\bibinfo {author} {\bibfnamefont {J.}~\bibnamefont
  {He}}, \bibinfo {author} {\bibfnamefont {Y.-H.}\ \bibnamefont {Zong}},
  \bibinfo {author} {\bibfnamefont {S.-P.}\ \bibnamefont {Kou}}, \bibinfo
  {author} {\bibfnamefont {Y.}~\bibnamefont {Liang}}, \ and\ \bibinfo {author}
  {\bibfnamefont {S.}~\bibnamefont {Feng}},\ }\bibfield  {title} {\enquote
  {\bibinfo {title} {Topological spin density waves in the {H}ubbard model on a
  honeycomb lattice},}\ }\href {\doibase 10.1103/PhysRevB.84.035127} {\bibfield
   {journal} {\bibinfo  {journal} {Phys. Rev. B}\ }\textbf {\bibinfo {volume}
  {84}},\ \bibinfo {pages} {035127} (\bibinfo {year} {2011})}\BibitemShut
  {NoStop}%
\bibitem [{\citenamefont {{He}}\ \emph {et~al.}(2012)\citenamefont {{He}},
  \citenamefont {{Liang}},\ and\ \citenamefont {{Kou}}}]{2012PhRvB..85t5107H}%
  \BibitemOpen
  \bibfield  {author} {\bibinfo {author} {\bibfnamefont {J.}~\bibnamefont
  {{He}}}, \bibinfo {author} {\bibfnamefont {Y.}~\bibnamefont {{Liang}}}, \
  and\ \bibinfo {author} {\bibfnamefont {S.-P.}\ \bibnamefont {{Kou}}},\
  }\bibfield  {title} {\enquote {\bibinfo {title} {{Composite spin liquid in a
  correlated topological insulator: Spin liquid without spin-charge
  separation}},}\ }\href {\doibase 10.1103/PhysRevB.85.205107} {\bibfield
  {journal} {\bibinfo  {journal} {\prb}\ }\textbf {\bibinfo {volume} {85}},\
  \bibinfo {eid} {205107} (\bibinfo {year} {2012})}\BibitemShut {NoStop}%
\bibitem [{\citenamefont {Zhu}\ \emph {et~al.}(2014)\citenamefont {Zhu},
  \citenamefont {He}, \citenamefont {Zang}, \citenamefont {Liang},\ and\
  \citenamefont {Kou}}]{0953-8984-26-17-175601}%
  \BibitemOpen
  \bibfield  {author} {\bibinfo {author} {\bibfnamefont {Y.-X.}\ \bibnamefont
  {Zhu}}, \bibinfo {author} {\bibfnamefont {J.}~\bibnamefont {He}}, \bibinfo
  {author} {\bibfnamefont {C.-L.}\ \bibnamefont {Zang}}, \bibinfo {author}
  {\bibfnamefont {Y.}~\bibnamefont {Liang}}, \ and\ \bibinfo {author}
  {\bibfnamefont {S.-P.}\ \bibnamefont {Kou}},\ }\bibfield  {title} {\enquote
  {\bibinfo {title} {Magnetic topological insulators at finite temperature},}\
  }\href {http://stacks.iop.org/0953-8984/26/i=17/a=175601} {\bibfield
  {journal} {\bibinfo  {journal} {Journal of Physics: Condensed Matter}\
  }\textbf {\bibinfo {volume} {26}},\ \bibinfo {pages} {175601} (\bibinfo
  {year} {2014})}\BibitemShut {NoStop}%
\bibitem [{\citenamefont {Wu}\ \emph {et~al.}(2015{\natexlab{a}})\citenamefont
  {Wu}, \citenamefont {Li},\ and\ \citenamefont {Kou}}]{Wu2015}%
  \BibitemOpen
  \bibfield  {author} {\bibinfo {author} {\bibfnamefont {Y.-J.}\ \bibnamefont
  {Wu}}, \bibinfo {author} {\bibfnamefont {N.}~\bibnamefont {Li}}, \ and\
  \bibinfo {author} {\bibfnamefont {S.-P.}\ \bibnamefont {Kou}},\ }\bibfield
  {title} {\enquote {\bibinfo {title} {Chiral topological superfluids in the
  attractive {H}aldane-{H}ubbard model with opposite {Z}eeman energy at two
  sublattice sites},}\ }\href@noop {} {\bibfield  {journal} {\bibinfo
  {journal} {The European Physical Journal B}\ }\textbf {\bibinfo {volume}
  {88}},\ \bibinfo {eid} {255} (\bibinfo {year}
  {2015}{\natexlab{a}})}\BibitemShut {NoStop}%
\bibitem [{\citenamefont {{Prychynenko}}\ and\ \citenamefont
  {{Huber}}(2016)}]{2016PhyB..481...53P}%
  \BibitemOpen
  \bibfield  {author} {\bibinfo {author} {\bibfnamefont {D.}~\bibnamefont
  {{Prychynenko}}}\ and\ \bibinfo {author} {\bibfnamefont {S.}~\bibnamefont
  {{Huber}}},\ }\bibfield  {title} {\enquote {\bibinfo {title} {{Z$_{2}$
  slave-spin theory of a strongly correlated {C}hern insulator}},}\ }\href
  {\doibase 10.1016/j.physb.2015.10.027} {\bibfield  {journal} {\bibinfo
  {journal} {Physica B Condensed Matter}\ }\textbf {\bibinfo {volume} {481}},\
  \bibinfo {pages} {53--58} (\bibinfo {year} {2016})}\BibitemShut {NoStop}%
\bibitem [{\citenamefont {Hung}\ \emph {et~al.}(2014)\citenamefont {Hung},
  \citenamefont {Chua}, \citenamefont {Wang},\ and\ \citenamefont
  {Fiete}}]{PhysRevB.89.235104}%
  \BibitemOpen
  \bibfield  {author} {\bibinfo {author} {\bibfnamefont {H.-H.}\ \bibnamefont
  {Hung}}, \bibinfo {author} {\bibfnamefont {V.}~\bibnamefont {Chua}}, \bibinfo
  {author} {\bibfnamefont {L.}~\bibnamefont {Wang}}, \ and\ \bibinfo {author}
  {\bibfnamefont {G.~A.}\ \bibnamefont {Fiete}},\ }\bibfield  {title} {\enquote
  {\bibinfo {title} {Interaction effects on topological phase transitions via
  numerically exact quantum {M}onte {C}arlo calculations},}\ }\href {\doibase
  10.1103/PhysRevB.89.235104} {\bibfield  {journal} {\bibinfo  {journal} {Phys.
  Rev. B}\ }\textbf {\bibinfo {volume} {89}},\ \bibinfo {pages} {235104}
  (\bibinfo {year} {2014})}\BibitemShut {NoStop}%
\bibitem [{\citenamefont {Zong}\ \emph {et~al.}(2013)\citenamefont {Zong},
  \citenamefont {He},\ and\ \citenamefont {Kou}}]{zong2013quantum}%
  \BibitemOpen
  \bibfield  {author} {\bibinfo {author} {\bibfnamefont {Y.-H.}\ \bibnamefont
  {Zong}}, \bibinfo {author} {\bibfnamefont {J.}~\bibnamefont {He}}, \ and\
  \bibinfo {author} {\bibfnamefont {S.-P.}\ \bibnamefont {Kou}},\ }\bibfield
  {title} {\enquote {\bibinfo {title} {Quantum spin liquid in interacting
  {K}ane-{M}ele model with staggered on-site potential},}\ }\href@noop {}
  {\bibfield  {journal} {\bibinfo  {journal} {The European Physical Journal B}\
  }\textbf {\bibinfo {volume} {86}},\ \bibinfo {pages} {1--10} (\bibinfo {year}
  {2013})}\BibitemShut {NoStop}%
\bibitem [{\citenamefont {Pesin}\ and\ \citenamefont
  {Balents}(2010)}]{2010NatPh...6..376P}%
  \BibitemOpen
  \bibfield  {author} {\bibinfo {author} {\bibfnamefont {D.}~\bibnamefont
  {Pesin}}\ and\ \bibinfo {author} {\bibfnamefont {L.}~\bibnamefont
  {Balents}},\ }\bibfield  {title} {\enquote {\bibinfo {title} {{M}ott physics
  and band topology in materials with strong spin-orbit interaction},}\ }\href
  {\doibase 10.1038/nphys1606} {\bibfield  {journal} {\bibinfo  {journal}
  {Nature Physics}\ }\textbf {\bibinfo {volume} {6}},\ \bibinfo {pages}
  {376--381} (\bibinfo {year} {2010})}\BibitemShut {NoStop}%
\bibitem [{\citenamefont {Hohenadler}\ \emph {et~al.}(2012)\citenamefont
  {Hohenadler}, \citenamefont {Meng}, \citenamefont {Lang}, \citenamefont
  {Wessel}, \citenamefont {Muramatsu},\ and\ \citenamefont
  {Assaad}}]{PhysRevB.85.115132}%
  \BibitemOpen
  \bibfield  {author} {\bibinfo {author} {\bibfnamefont {M.}~\bibnamefont
  {Hohenadler}}, \bibinfo {author} {\bibfnamefont {Z.~Y.}\ \bibnamefont
  {Meng}}, \bibinfo {author} {\bibfnamefont {T.~C.}\ \bibnamefont {Lang}},
  \bibinfo {author} {\bibfnamefont {S.}~\bibnamefont {Wessel}}, \bibinfo
  {author} {\bibfnamefont {A.}~\bibnamefont {Muramatsu}}, \ and\ \bibinfo
  {author} {\bibfnamefont {F.~F.}\ \bibnamefont {Assaad}},\ }\bibfield  {title}
  {\enquote {\bibinfo {title} {Quantum phase transitions in the
  {K}ane-{M}ele-{H}ubbard model},}\ }\href {\doibase
  10.1103/PhysRevB.85.115132} {\bibfield  {journal} {\bibinfo  {journal} {Phys.
  Rev. B}\ }\textbf {\bibinfo {volume} {85}},\ \bibinfo {pages} {115132}
  (\bibinfo {year} {2012})}\BibitemShut {NoStop}%
\bibitem [{\citenamefont {Chen}\ \emph {et~al.}(2015)\citenamefont {Chen},
  \citenamefont {Hung}, \citenamefont {Su}, \citenamefont {Fiete},\ and\
  \citenamefont {Ting}}]{PhysRevB.91.045122}%
  \BibitemOpen
  \bibfield  {author} {\bibinfo {author} {\bibfnamefont {Y.-H.}\ \bibnamefont
  {Chen}}, \bibinfo {author} {\bibfnamefont {H.-H.}\ \bibnamefont {Hung}},
  \bibinfo {author} {\bibfnamefont {G.}~\bibnamefont {Su}}, \bibinfo {author}
  {\bibfnamefont {G.~A.}\ \bibnamefont {Fiete}}, \ and\ \bibinfo {author}
  {\bibfnamefont {C.~S.}\ \bibnamefont {Ting}},\ }\bibfield  {title} {\enquote
  {\bibinfo {title} {Cellular dynamical mean-field theory study of an
  interacting topological honeycomb lattice model at finite temperature},}\
  }\href {\doibase 10.1103/PhysRevB.91.045122} {\bibfield  {journal} {\bibinfo
  {journal} {Phys. Rev. B}\ }\textbf {\bibinfo {volume} {91}},\ \bibinfo
  {pages} {045122} (\bibinfo {year} {2015})}\BibitemShut {NoStop}%
\bibitem [{\citenamefont {{K}ane}\ and\ \citenamefont
  {{M}ele}(2005)}]{PhysRevLett.95.226801}%
  \BibitemOpen
  \bibfield  {author} {\bibinfo {author} {\bibfnamefont {C.~L.}\ \bibnamefont
  {{K}ane}}\ and\ \bibinfo {author} {\bibfnamefont {E.~J.}\ \bibnamefont
  {{M}ele}},\ }\bibfield  {title} {\enquote {\bibinfo {title} {Quantum spin
  hall effect in graphene},}\ }\href {\doibase 10.1103/PhysRevLett.95.226801}
  {\bibfield  {journal} {\bibinfo  {journal} {Phys. Rev. Lett.}\ }\textbf
  {\bibinfo {volume} {95}},\ \bibinfo {pages} {226801} (\bibinfo {year}
  {2005})}\BibitemShut {NoStop}%
\bibitem [{\citenamefont {Liu}\ \emph {et~al.}(2016)\citenamefont {Liu},
  \citenamefont {Liu}, \citenamefont {Law}, \citenamefont {Liu},\ and\
  \citenamefont {Ng}}]{1367-2630-18-3-035004}%
  \BibitemOpen
  \bibfield  {author} {\bibinfo {author} {\bibfnamefont {X.-J.}\ \bibnamefont
  {Liu}}, \bibinfo {author} {\bibfnamefont {Z.-X.}\ \bibnamefont {Liu}},
  \bibinfo {author} {\bibfnamefont {K.~T.}\ \bibnamefont {Law}}, \bibinfo
  {author} {\bibfnamefont {W.~V.}\ \bibnamefont {Liu}}, \ and\ \bibinfo
  {author} {\bibfnamefont {T.~K.}\ \bibnamefont {Ng}},\ }\bibfield  {title}
  {\enquote {\bibinfo {title} {Chiral topological orders in an optical raman
  lattice},}\ }\href {http://stacks.iop.org/1367-2630/18/i=3/a=035004}
  {\bibfield  {journal} {\bibinfo  {journal} {New Journal of Physics}\ }\textbf
  {\bibinfo {volume} {18}},\ \bibinfo {pages} {035004} (\bibinfo {year}
  {2016})}\BibitemShut {NoStop}%
\bibitem [{\citenamefont {Hickey}\ \emph
  {et~al.}(2015{\natexlab{a}})\citenamefont {Hickey}, \citenamefont {Rath},\
  and\ \citenamefont {Paramekanti}}]{2015PhRvB..91m4414H}%
  \BibitemOpen
  \bibfield  {author} {\bibinfo {author} {\bibfnamefont {C.}~\bibnamefont
  {Hickey}}, \bibinfo {author} {\bibfnamefont {P.}~\bibnamefont {Rath}}, \ and\
  \bibinfo {author} {\bibfnamefont {A.}~\bibnamefont {Paramekanti}},\
  }\bibfield  {title} {\enquote {\bibinfo {title} {Competing chiral orders in
  the topological {H}aldane-{H}ubbard model of spin-1/2 fermions and bosons},}\
  }\href {\doibase 10.1103/PhysRevB.91.134414} {\bibfield  {journal} {\bibinfo
  {journal} {\prb}\ }\textbf {\bibinfo {volume} {91}},\ \bibinfo {eid} {134414}
  (\bibinfo {year} {2015}{\natexlab{a}})}\BibitemShut {NoStop}%
\bibitem [{\citenamefont {Hickey}\ \emph
  {et~al.}(2015{\natexlab{b}})\citenamefont {Hickey}, \citenamefont {Cincio},
  \citenamefont {Papi{\'c}},\ and\ \citenamefont
  {Paramekanti}}]{2015arXiv150908461H}%
  \BibitemOpen
  \bibfield  {author} {\bibinfo {author} {\bibfnamefont {C.}~\bibnamefont
  {Hickey}}, \bibinfo {author} {\bibfnamefont {L.}~\bibnamefont {Cincio}},
  \bibinfo {author} {\bibfnamefont {Z.}~\bibnamefont {Papi{\'c}}}, \ and\
  \bibinfo {author} {\bibfnamefont {A.}~\bibnamefont {Paramekanti}},\
  }\bibfield  {title} {\enquote {\bibinfo {title} {{H}aldane-{H}ubbard {M}ott
  insulator: From tetrahedral spin crystal to chiral spin liquid},}\
  }\href@noop {} {\bibfield  {journal} {\bibinfo  {journal} {ArXiv e-prints}\ }
  (\bibinfo {year} {2015}{\natexlab{b}})},\ \Eprint
  {http://arxiv.org/abs/1509.08461} {arXiv:1509.08461 [cond-mat.str-el]}
  \BibitemShut {NoStop}%
\bibitem [{\citenamefont {Zheng}\ \emph {et~al.}(2015)\citenamefont {Zheng},
  \citenamefont {Shen}, \citenamefont {Wang},\ and\ \citenamefont
  {Zhai}}]{PhysRevB.91.161107}%
  \BibitemOpen
  \bibfield  {author} {\bibinfo {author} {\bibfnamefont {W.}~\bibnamefont
  {Zheng}}, \bibinfo {author} {\bibfnamefont {H.}~\bibnamefont {Shen}},
  \bibinfo {author} {\bibfnamefont {Z.}~\bibnamefont {Wang}}, \ and\ \bibinfo
  {author} {\bibfnamefont {H.}~\bibnamefont {Zhai}},\ }\bibfield  {title}
  {\enquote {\bibinfo {title} {Magnetic-order-driven topological transition in
  the {H}aldane-{H}ubbard model},}\ }\href {\doibase
  10.1103/PhysRevB.91.161107} {\bibfield  {journal} {\bibinfo  {journal} {Phys.
  Rev. B}\ }\textbf {\bibinfo {volume} {91}},\ \bibinfo {pages} {161107}
  (\bibinfo {year} {2015})}\BibitemShut {NoStop}%
\bibitem [{\citenamefont {Gu}\ \emph {et~al.}(2015)\citenamefont {Gu},
  \citenamefont {Li},\ and\ \citenamefont {Li}}]{2015arXiv151205118G}%
  \BibitemOpen
  \bibfield  {author} {\bibinfo {author} {\bibfnamefont {Z.~L.}\ \bibnamefont
  {Gu}}, \bibinfo {author} {\bibfnamefont {K.}~\bibnamefont {Li}}, \ and\
  \bibinfo {author} {\bibfnamefont {J.~X.}\ \bibnamefont {Li}},\ }\bibfield
  {title} {\enquote {\bibinfo {title} {Topological phase transitions and
  topological {M}ott insulator in {H}aldane-{H}ubbard model},}\ }\href@noop {}
  {\bibfield  {journal} {\bibinfo  {journal} {ArXiv e-prints}\ } (\bibinfo
  {year} {2015})},\ \Eprint {http://arxiv.org/abs/1512.05118} {arXiv:1512.05118
  [cond-mat.str-el]} \BibitemShut {NoStop}%
\bibitem [{\citenamefont {Wu}\ \emph {et~al.}(2015{\natexlab{b}})\citenamefont
  {Wu}, \citenamefont {Faye}, \citenamefont {S{\'{e}}n{\'{e}}chal},\ and\
  \citenamefont {Maciejko}}]{2015arXiv151204498W}%
  \BibitemOpen
  \bibfield  {author} {\bibinfo {author} {\bibfnamefont {J.}~\bibnamefont
  {Wu}}, \bibinfo {author} {\bibfnamefont {J.~P.~L.}\ \bibnamefont {Faye}},
  \bibinfo {author} {\bibfnamefont {D.}~\bibnamefont {S{\'{e}}n{\'{e}}chal}}, \
  and\ \bibinfo {author} {\bibfnamefont {J.}~\bibnamefont {Maciejko}},\
  }\bibfield  {title} {\enquote {\bibinfo {title} {A quantum cluster approach
  to the spinful {H}aldane-{H}ubbard model},}\ }\href@noop {} {\bibfield
  {journal} {\bibinfo  {journal} {ArXiv e-prints}\ } (\bibinfo {year}
  {2015}{\natexlab{b}})},\ \Eprint {http://arxiv.org/abs/1512.04498}
  {arXiv:1512.04498 [cond-mat.str-el]} \BibitemShut {NoStop}%
\bibitem [{\citenamefont {Zhang}\ \emph {et~al.}(2015)\citenamefont {Zhang},
  \citenamefont {Xu},\ and\ \citenamefont {Zhang}}]{2015arXiv151103833Z}%
  \BibitemOpen
  \bibfield  {author} {\bibinfo {author} {\bibfnamefont {Y.~C.}\ \bibnamefont
  {Zhang}}, \bibinfo {author} {\bibfnamefont {Z.}~\bibnamefont {Xu}}, \ and\
  \bibinfo {author} {\bibfnamefont {S.}~\bibnamefont {Zhang}},\ }\bibfield
  {title} {\enquote {\bibinfo {title} {Topological superfluids and
  {{BEC}}-{{BCS}} crossover in attractive {H}aldane-{H}ubbard model},}\
  }\href@noop {} {\bibfield  {journal} {\bibinfo  {journal} {ArXiv e-prints}\ }
  (\bibinfo {year} {2015})},\ \Eprint {http://arxiv.org/abs/1511.03833}
  {arXiv:1511.03833 [cond-mat.quant-gas]} \BibitemShut {NoStop}%
\bibitem [{\citenamefont {Georges}\ \emph {et~al.}(1996)\citenamefont
  {Georges}, \citenamefont {Kotliar}, \citenamefont {Krauth},\ and\
  \citenamefont {Rozenberg}}]{RevModPhys.68.}%
  \BibitemOpen
  \bibfield  {author} {\bibinfo {author} {\bibfnamefont {A.}~\bibnamefont
  {Georges}}, \bibinfo {author} {\bibfnamefont {G.}~\bibnamefont {Kotliar}},
  \bibinfo {author} {\bibfnamefont {W.}~\bibnamefont {Krauth}}, \ and\ \bibinfo
  {author} {\bibfnamefont {M.~J.}\ \bibnamefont {Rozenberg}},\ }\bibfield
  {title} {\enquote {\bibinfo {title} {Dynamical mean-field theory of strongly
  correlated fermion systems and the limit of infinite dimensions},}\ }\href
  {\doibase 10.1103/RevModPhys.68.13} {\bibfield  {journal} {\bibinfo
  {journal} {Rev. Mod. Phys.}\ }\textbf {\bibinfo {volume} {68}},\ \bibinfo
  {pages} {13--125} (\bibinfo {year} {1996})}\BibitemShut {NoStop}%
\bibitem [{\citenamefont {Caffarel}\ and\ \citenamefont
  {Krauth}(1994)}]{PhysRevLett.72.1545}%
  \BibitemOpen
  \bibfield  {author} {\bibinfo {author} {\bibfnamefont {M.}~\bibnamefont
  {Caffarel}}\ and\ \bibinfo {author} {\bibfnamefont {W.}~\bibnamefont
  {Krauth}},\ }\bibfield  {title} {\enquote {\bibinfo {title} {Exact
  diagonalization approach to correlated fermions in infinite dimensions:
  {M}ott transition and superconductivity},}\ }\href {\doibase
  10.1103/PhysRevLett.72.1545} {\bibfield  {journal} {\bibinfo  {journal}
  {Phys. Rev. Lett.}\ }\textbf {\bibinfo {volume} {72}},\ \bibinfo {pages}
  {1545--1548} (\bibinfo {year} {1994})}\BibitemShut {NoStop}%
\bibitem [{\citenamefont {Maier}\ \emph {et~al.}(2005)\citenamefont {Maier},
  \citenamefont {Jarrell}, \citenamefont {Pruschke},\ and\ \citenamefont
  {Hettler}}]{RevModPhys.77.}%
  \BibitemOpen
  \bibfield  {author} {\bibinfo {author} {\bibfnamefont {T.}~\bibnamefont
  {Maier}}, \bibinfo {author} {\bibfnamefont {M.}~\bibnamefont {Jarrell}},
  \bibinfo {author} {\bibfnamefont {T.}~\bibnamefont {Pruschke}}, \ and\
  \bibinfo {author} {\bibfnamefont {M.~H.}\ \bibnamefont {Hettler}},\
  }\bibfield  {title} {\enquote {\bibinfo {title} {Quantum cluster theories},}\
  }\href {\doibase 10.1103/RevModPhys.77.1027} {\bibfield  {journal} {\bibinfo
  {journal} {Rev. Mod. Phys.}\ }\textbf {\bibinfo {volume} {77}},\ \bibinfo
  {pages} {1027--1080} (\bibinfo {year} {2005})}\BibitemShut {NoStop}%
\bibitem [{\citenamefont {Kotliar}\ \emph {et~al.}(2001)\citenamefont
  {Kotliar}, \citenamefont {Savrasov}, \citenamefont {P{\'{a}}lsson},\ and\
  \citenamefont {Biroli}}]{PhysRevLett.87.186401}%
  \BibitemOpen
  \bibfield  {author} {\bibinfo {author} {\bibfnamefont {G.}~\bibnamefont
  {Kotliar}}, \bibinfo {author} {\bibfnamefont {S.~Y.}\ \bibnamefont
  {Savrasov}}, \bibinfo {author} {\bibfnamefont {G.}~\bibnamefont
  {P{\'{a}}lsson}}, \ and\ \bibinfo {author} {\bibfnamefont {G.}~\bibnamefont
  {Biroli}},\ }\bibfield  {title} {\enquote {\bibinfo {title} {Cellular
  dynamical mean field approach to strongly correlated systems},}\ }\href
  {\doibase 10.1103/PhysRevLett.87.186401} {\bibfield  {journal} {\bibinfo
  {journal} {Phys. Rev. Lett.}\ }\textbf {\bibinfo {volume} {87}},\ \bibinfo
  {pages} {186401} (\bibinfo {year} {2001})}\BibitemShut {NoStop}%
\bibitem [{\citenamefont {Rubtsov}\ \emph {et~al.}(2005)\citenamefont
  {Rubtsov}, \citenamefont {Savkin},\ and\ \citenamefont
  {Lichtenstein}}]{PhysRevB.72.035122}%
  \BibitemOpen
  \bibfield  {author} {\bibinfo {author} {\bibfnamefont {A.~N.}\ \bibnamefont
  {Rubtsov}}, \bibinfo {author} {\bibfnamefont {V.~V.}\ \bibnamefont {Savkin}},
  \ and\ \bibinfo {author} {\bibfnamefont {A.~I.}\ \bibnamefont
  {Lichtenstein}},\ }\bibfield  {title} {\enquote {\bibinfo {title}
  {Continuous-time quantum {M}onte {C}arlo method for fermions},}\ }\href
  {\doibase 10.1103/PhysRevB.72.035122} {\bibfield  {journal} {\bibinfo
  {journal} {Phys. Rev. B}\ }\textbf {\bibinfo {volume} {72}},\ \bibinfo
  {pages} {035122} (\bibinfo {year} {2005})}\BibitemShut {NoStop}%
\bibitem [{\citenamefont {Gull}\ \emph {et~al.}(2011)\citenamefont {Gull},
  \citenamefont {Millis}, \citenamefont {Lichtenstein}, \citenamefont
  {Rubtsov}, \citenamefont {Troyer},\ and\ \citenamefont
  {Werner}}]{ContinuousTime}%
  \BibitemOpen
  \bibfield  {author} {\bibinfo {author} {\bibfnamefont {E.}~\bibnamefont
  {Gull}}, \bibinfo {author} {\bibfnamefont {A.~J.}\ \bibnamefont {Millis}},
  \bibinfo {author} {\bibfnamefont {A.~I.}\ \bibnamefont {Lichtenstein}},
  \bibinfo {author} {\bibfnamefont {A.~N.}\ \bibnamefont {Rubtsov}}, \bibinfo
  {author} {\bibfnamefont {M.}~\bibnamefont {Troyer}}, \ and\ \bibinfo {author}
  {\bibfnamefont {P.}~\bibnamefont {Werner}},\ }\bibfield  {title} {\enquote
  {\bibinfo {title} {Continuous-time {M}onte {C}arlo methods for quantum
  impurity models},}\ }\href {\doibase 10.1103/RevModPhys.83.349} {\bibfield
  {journal} {\bibinfo  {journal} {Reviews of Modern Physics}\ }\textbf
  {\bibinfo {volume} {83}},\ \bibinfo {pages} {349--404} (\bibinfo {year}
  {2011})}\BibitemShut {NoStop}%
\bibitem [{\citenamefont {Wang}\ and\ \citenamefont
  {Zhang}(2012)}]{PhysRevX.2.031008}%
  \BibitemOpen
  \bibfield  {author} {\bibinfo {author} {\bibfnamefont {Z.}~\bibnamefont
  {Wang}}\ and\ \bibinfo {author} {\bibfnamefont {S.-C.}\ \bibnamefont
  {Zhang}},\ }\bibfield  {title} {\enquote {\bibinfo {title} {Simplified
  topological invariants for interacting insulators},}\ }\href {\doibase
  10.1103/PhysRevX.2.031008} {\bibfield  {journal} {\bibinfo  {journal} {Phys.
  Rev. X}\ }\textbf {\bibinfo {volume} {2}},\ \bibinfo {pages} {031008}
  (\bibinfo {year} {2012})}\BibitemShut {NoStop}%
\bibitem [{\citenamefont {Wang}\ and\ \citenamefont
  {Yan}(2013)}]{0953_8984_25_15_155601}%
  \BibitemOpen
  \bibfield  {author} {\bibinfo {author} {\bibfnamefont {Z.}~\bibnamefont
  {Wang}}\ and\ \bibinfo {author} {\bibfnamefont {B.}~\bibnamefont {Yan}},\
  }\bibfield  {title} {\enquote {\bibinfo {title} {Topological {H}amiltonian as
  an exact tool for topological invariants},}\ }\href
  {http://stacks.iop.org/0953-8984/25/i=15/a=155601} {\bibfield  {journal}
  {\bibinfo  {journal} {Journal of Physics: Condensed Matter}\ }\textbf
  {\bibinfo {volume} {25}},\ \bibinfo {pages} {155601} (\bibinfo {year}
  {2013})}\BibitemShut {NoStop}%
\bibitem [{\citenamefont {Witczak-Krempa}\ \emph {et~al.}(2014)\citenamefont
  {Witczak-Krempa}, \citenamefont {Knap},\ and\ \citenamefont
  {Abanin}}]{PhysRevLett.113.136402}%
  \BibitemOpen
  \bibfield  {author} {\bibinfo {author} {\bibfnamefont {W.}~\bibnamefont
  {Witczak-Krempa}}, \bibinfo {author} {\bibfnamefont {M.}~\bibnamefont
  {Knap}}, \ and\ \bibinfo {author} {\bibfnamefont {D.}~\bibnamefont
  {Abanin}},\ }\bibfield  {title} {\enquote {\bibinfo {title} {Interacting
  {W}eyl semimetals: Characterization via the topological {H}amiltonian and its
  breakdown},}\ }\href {\doibase 10.1103/PhysRevLett.113.136402} {\bibfield
  {journal} {\bibinfo  {journal} {Phys. Rev. Lett.}\ }\textbf {\bibinfo
  {volume} {113}},\ \bibinfo {pages} {136402} (\bibinfo {year}
  {2014})}\BibitemShut {NoStop}%
\bibitem [{\citenamefont {Fukui}\ \emph {et~al.}(2005)\citenamefont {Fukui},
  \citenamefont {Hatsugai},\ and\ \citenamefont
  {Suzuki}}]{FukuiTakahiroHatsugai2005}%
  \BibitemOpen
  \bibfield  {author} {\bibinfo {author} {\bibfnamefont {T.}~\bibnamefont
  {Fukui}}, \bibinfo {author} {\bibfnamefont {Y.}~\bibnamefont {Hatsugai}}, \
  and\ \bibinfo {author} {\bibfnamefont {H.}~\bibnamefont {Suzuki}},\
  }\bibfield  {title} {\enquote {\bibinfo {title} {{C}hern numbers in
  discretized brillouin zone: Efficient method of computing (spin) hall
  conductances},}\ }\href {\doibase 10.1143/JPSJ.74.1674} {\bibfield  {journal}
  {\bibinfo  {journal} {Journal of the Physical Society of Japan}\ }\textbf
  {\bibinfo {volume} {74}},\ \bibinfo {pages} {1674--1677} (\bibinfo {year}
  {2005})}\BibitemShut {NoStop}%
\bibitem [{\citenamefont {Niu}\ \emph {et~al.}(1985)\citenamefont {Niu},
  \citenamefont {Thouless},\ and\ \citenamefont {Wu}}]{PhysRevB.31.3372}%
  \BibitemOpen
  \bibfield  {author} {\bibinfo {author} {\bibfnamefont {Q.}~\bibnamefont
  {Niu}}, \bibinfo {author} {\bibfnamefont {D.~J.}\ \bibnamefont {Thouless}}, \
  and\ \bibinfo {author} {\bibfnamefont {Y.-S.}\ \bibnamefont {Wu}},\
  }\bibfield  {title} {\enquote {\bibinfo {title} {Quantized hall conductance
  as a topological invariant},}\ }\href {\doibase 10.1103/PhysRevB.31.3372}
  {\bibfield  {journal} {\bibinfo  {journal} {Phys. Rev. B}\ }\textbf {\bibinfo
  {volume} {31}},\ \bibinfo {pages} {3372--3377} (\bibinfo {year}
  {1985})}\BibitemShut {NoStop}%
\bibitem [{\citenamefont {Resta}(2011)}]{Resta_2011}%
  \BibitemOpen
  \bibfield  {author} {\bibinfo {author} {\bibfnamefont {R.}~\bibnamefont
  {Resta}},\ }\bibfield  {title} {\enquote {\bibinfo {title} {The insulating
  state of matter: a geometrical theory},}\ }\href {\doibase
  10.1140/epjb/e2010-10874-4} {\bibfield  {journal} {\bibinfo  {journal} {The
  European Physical Journal B}\ }\textbf {\bibinfo {volume} {79}},\ \bibinfo
  {pages} {121--137} (\bibinfo {year} {2011})}\BibitemShut {NoStop}%
\bibitem [{\citenamefont {Varney}\ \emph {et~al.}(2011)\citenamefont {Varney},
  \citenamefont {Sun}, \citenamefont {Rigol},\ and\ \citenamefont
  {Galitski}}]{PhysRevB.84.241105}%
  \BibitemOpen
  \bibfield  {author} {\bibinfo {author} {\bibfnamefont {C.~N.}\ \bibnamefont
  {Varney}}, \bibinfo {author} {\bibfnamefont {K.}~\bibnamefont {Sun}},
  \bibinfo {author} {\bibfnamefont {M.}~\bibnamefont {Rigol}}, \ and\ \bibinfo
  {author} {\bibfnamefont {V.}~\bibnamefont {Galitski}},\ }\bibfield  {title}
  {\enquote {\bibinfo {title} {Topological phase transitions for interacting
  finite systems},}\ }\href {\doibase 10.1103/PhysRevB.84.241105} {\bibfield
  {journal} {\bibinfo  {journal} {Phys. Rev. B}\ }\textbf {\bibinfo {volume}
  {84}},\ \bibinfo {pages} {241105} (\bibinfo {year} {2011})}\BibitemShut
  {NoStop}%
\bibitem [{\citenamefont {T{\"{o}}rm{\"{a}}}(2015)}]{PaivinSpektroskopia2015}%
  \BibitemOpen
  \bibfield  {author} {\bibinfo {author} {\bibfnamefont {P.}~\bibnamefont
  {T{\"{o}}rm{\"{a}}}},\ }\bibfield  {title} {\enquote {\bibinfo {title}
  {Spectroscopies {--} theory},}\ }in\ \href@noop {} {\emph {\bibinfo
  {booktitle} {Quantum Gas Experiments {--} Exploring Many-Body States}}},\
  \bibinfo {editor} {edited by\ \bibinfo {editor} {\bibfnamefont
  {P.}~\bibnamefont {T{\"{o}}rm{\"{a}}}}\ and\ \bibinfo {editor} {\bibfnamefont
  {K.}~\bibnamefont {Sengstock}}}\ (\bibinfo  {publisher} {Imperial College
  Press},\ \bibinfo {address} {London},\ \bibinfo {year} {2015})\BibitemShut
  {NoStop}%
\bibitem [{\citenamefont {Liebsch}\ and\ \citenamefont
  {Wu}(2013)}]{PhysRevB.87.205127}%
  \BibitemOpen
  \bibfield  {author} {\bibinfo {author} {\bibfnamefont {A.}~\bibnamefont
  {Liebsch}}\ and\ \bibinfo {author} {\bibfnamefont {W.}~\bibnamefont {Wu}},\
  }\bibfield  {title} {\enquote {\bibinfo {title} {Coulomb correlations in the
  honeycomb lattice: Role of translation symmetry},}\ }\href {\doibase
  10.1103/PhysRevB.87.205127} {\bibfield  {journal} {\bibinfo  {journal} {Phys.
  Rev. B}\ }\textbf {\bibinfo {volume} {87}},\ \bibinfo {pages} {205127}
  (\bibinfo {year} {2013})}\BibitemShut {NoStop}%
\bibitem [{\citenamefont {Sorella}\ \emph {et~al.}(2012)\citenamefont
  {Sorella}, \citenamefont {Otsuka},\ and\ \citenamefont
  {Yunoki}}]{2012NatSR...2E.992S}%
  \BibitemOpen
  \bibfield  {author} {\bibinfo {author} {\bibfnamefont {S.}~\bibnamefont
  {Sorella}}, \bibinfo {author} {\bibfnamefont {Y.}~\bibnamefont {Otsuka}}, \
  and\ \bibinfo {author} {\bibfnamefont {S.}~\bibnamefont {Yunoki}},\
  }\bibfield  {title} {\enquote {\bibinfo {title} {Absence of a spin liquid
  phase in the {H}ubbard model on the honeycomb lattice},}\ }\href {\doibase
  10.1038/srep00992} {\bibfield  {journal} {\bibinfo  {journal} {Scientific
  Reports}\ }\textbf {\bibinfo {volume} {2}},\ \bibinfo {eid} {992} (\bibinfo
  {year} {2012})}\BibitemShut {NoStop}%
\bibitem [{\citenamefont {He}\ and\ \citenamefont
  {Lu}(2012)}]{PhysRevB.86.045105}%
  \BibitemOpen
  \bibfield  {author} {\bibinfo {author} {\bibfnamefont {R.-Q.}\ \bibnamefont
  {He}}\ and\ \bibinfo {author} {\bibfnamefont {Z.-Y.}\ \bibnamefont {Lu}},\
  }\bibfield  {title} {\enquote {\bibinfo {title} {Cluster dynamical mean field
  theory of quantum phases on a honeycomb lattice},}\ }\href {\doibase
  10.1103/PhysRevB.86.045105} {\bibfield  {journal} {\bibinfo  {journal} {Phys.
  Rev. B}\ }\textbf {\bibinfo {volume} {86}},\ \bibinfo {pages} {045105}
  (\bibinfo {year} {2012})}\BibitemShut {NoStop}%
\bibitem [{\citenamefont {Cuccoli}\ \emph
  {et~al.}(2003{\natexlab{a}})\citenamefont {Cuccoli}, \citenamefont
  {Roscilde}, \citenamefont {Tognetti}, \citenamefont {Vaia},\ and\
  \citenamefont {Verrucchi}}]{PhysRevB.67.104414}%
  \BibitemOpen
  \bibfield  {author} {\bibinfo {author} {\bibfnamefont {A.}~\bibnamefont
  {Cuccoli}}, \bibinfo {author} {\bibfnamefont {T.}~\bibnamefont {Roscilde}},
  \bibinfo {author} {\bibfnamefont {V.}~\bibnamefont {Tognetti}}, \bibinfo
  {author} {\bibfnamefont {R.}~\bibnamefont {Vaia}}, \ and\ \bibinfo {author}
  {\bibfnamefont {P.}~\bibnamefont {Verrucchi}},\ }\bibfield  {title} {\enquote
  {\bibinfo {title} {Quantum {M}onte {C}arlo study of $s=\frac12$ weakly
  anisotropic antiferromagnets on the square lattice},}\ }\href {\doibase
  10.1103/PhysRevB.67.104414} {\bibfield  {journal} {\bibinfo  {journal} {Phys.
  Rev. B}\ }\textbf {\bibinfo {volume} {67}},\ \bibinfo {pages} {104414}
  (\bibinfo {year} {2003}{\natexlab{a}})}\BibitemShut {NoStop}%
\bibitem [{\citenamefont {Cuccoli}\ \emph
  {et~al.}(2003{\natexlab{b}})\citenamefont {Cuccoli}, \citenamefont
  {Roscilde}, \citenamefont {Vaia},\ and\ \citenamefont
  {Verrucchi}}]{PhysRevB.68.060402}%
  \BibitemOpen
  \bibfield  {author} {\bibinfo {author} {\bibfnamefont {A.}~\bibnamefont
  {Cuccoli}}, \bibinfo {author} {\bibfnamefont {T.}~\bibnamefont {Roscilde}},
  \bibinfo {author} {\bibfnamefont {R.}~\bibnamefont {Vaia}}, \ and\ \bibinfo
  {author} {\bibfnamefont {P.}~\bibnamefont {Verrucchi}},\ }\bibfield  {title}
  {\enquote {\bibinfo {title} {Field-induced $\mathrm{XY}$ behavior in the
  $s=\frac{1}{2}$ antiferromagnet on the square lattice},}\ }\href {\doibase
  10.1103/PhysRevB.68.060402} {\bibfield  {journal} {\bibinfo  {journal} {Phys.
  Rev. B}\ }\textbf {\bibinfo {volume} {68}},\ \bibinfo {pages} {060402}
  (\bibinfo {year} {2003}{\natexlab{b}})}\BibitemShut {NoStop}%
\bibitem [{\citenamefont {{He}}\ \emph {et~al.}(2011)\citenamefont {{He}},
  \citenamefont {{Kou}}, \citenamefont {{Liang}},\ and\ \citenamefont
  {{Feng}}}]{2011PhRvB..83t5116H}%
  \BibitemOpen
  \bibfield  {author} {\bibinfo {author} {\bibfnamefont {J.}~\bibnamefont
  {{He}}}, \bibinfo {author} {\bibfnamefont {S.-P.}\ \bibnamefont {{Kou}}},
  \bibinfo {author} {\bibfnamefont {Y.}~\bibnamefont {{Liang}}}, \ and\
  \bibinfo {author} {\bibfnamefont {S.}~\bibnamefont {{Feng}}},\ }\bibfield
  {title} {\enquote {\bibinfo {title} {{Chiral spin liquid in a correlated
  topological insulator}},}\ }\href {\doibase 10.1103/PhysRevB.83.205116}
  {\bibfield  {journal} {\bibinfo  {journal} {\prb}\ }\textbf {\bibinfo
  {volume} {83}},\ \bibinfo {eid} {205116} (\bibinfo {year}
  {2011})}\BibitemShut {NoStop}%
\end{thebibliography}%

\end{document}